\documentclass[fleqn,12pt,twoside]{article}
\usepackage[headings]{espcrc1}

\readRCS
$Id: espcrc1.tex,v 1.2 2004/02/24 11:22:11 spepping Exp $
\usepackage{graphicx}
\usepackage[figuresright]{rotating}

\newcommand{\AmS}{{\protect\the\textfont2
  A\kern-.1667em\lower.5ex\hbox{M}\kern-.125emS}}

\title{Neutrinos and Symmetries}

\author{A.B. Balantekin\address[MCSD]{Department of Physics, University of Wisconsin, 
        Madison, WI 53706,  USA}}%
        
\runtitle{Neutrinos and Symmetries}
\runauthor{A.B. Balantekin}

\begin{document}

\maketitle

\begin{abstract}
Three facets of symmetries in neutrino physics are briefly reviewed: 
i) The SO(5) symmetry of the neutrino mass and and its connection to the 
see-saw mechanism; ii)  SU(N$_{\rm flavor}$) symmetries of dense, 
self-interacting neutrino gases in astrophysical settings; iii) The neutrino 
mixing angle $\theta_{13}$ and 
possible CP-violation in the neutrino sector. 
\end{abstract}

\section{SO(5) SYMMETRY OF THE NEUTRINO MASS}

One of the most exciting discoveries in the physics during the last two decades 
was that neutrinos have mass. This was achieved through the observation of 
neutrino oscillations. There is no neutrino mass term in the Standard Model. 
However, using the Standard Model degrees of freedom one can parameterize the 
neutrino mass by a dimension-five operator: $X_{\alpha \beta} H H 
\overline{\nu_{L \alpha}^C} \nu_{L \beta} / \Lambda$. This term is not renormalizable. 
Furthermore it is the only dimension-five operator one can write using the Standard Model 
degrees of freedom. Hence the neutrino mass is, in some sense, the most accessible 
new physics beyond the Standard Model. One can also write down another mass term:  
$\overline{\nu_R^C}  \nu_R$, which is permitted by the weak-isospin 
invariance of the Standard Model.  Such Majorana mass terms  
violate lepton number conservation since they imply that neutrinos are their own 
antiparticles.  

Back in 1957, Pauli and G\"ursey considered a particle-antiparticle symmetry, realized 
via the transformation 
\begin{equation}
\Psi \rightarrow a \Psi + b \gamma_5 \Psi^C, \>\>\> |a|^2 + |b|^2 =1. 
\end{equation}
It is easy to see that , under such a transformation, a pure Dirac mass terms would transform 
into a mixture of Dirac and Majorana mass terms. Indeed it is easy to show that the 
"charge" (as opposed to "current") operators 
\begin{eqnarray}
D_+ &=& \frac{1}{2}  \int d^3{\bf x} \overline{\Psi_L} \Psi_R,  \\
A_+ &=& \int d^3{\bf x} \left[ - \Psi_L^T {\cal C} \gamma_0 \Psi_R \right], \\
L_+ &=&  \frac{1}{2}  \int d^3{\bf x}  \left( \overline{\Psi}_L \Psi_L^C \right), \>\>  R_+ = \frac{1}{2}  \int d^3{\bf x} \left( \overline{\Psi^C_R} \Psi_R \right), 
\end{eqnarray}
their complex conjugates, and the operators
\begin{equation}
L_0 =  \frac{1}{4}  \int d^3{\bf x} \left( \Psi^{\dagger}_L \Psi_L - \Psi_L \Psi^{\dagger}_L \right), \>\>\>
R_0 =  \frac{1}{4}  \int d^3{\bf x} \left( \Psi_R \Psi^{\dagger}_R - \Psi^{\dagger}_R \Psi_R \right) 
\end{equation}
form an SO(5) algebra \cite{Balantekin:2000qt}. The operators $A_+, A_-$ and 
$A_0 = R_0 -L_0$ form an SU(2) subalgebra that generates the Pauli-G\"ursey transformation 
(SU(2)$_{\rm PG}$).  
The most general neutrino mass Hamiltonian sits in the $SO(5)/SU(2)_L \times SU(2)_R 
\times  U(1)_{L_0+R_0}$ coset and can be diagonalized by a SU(2)$_{\rm PG}$ rotation, 
producing the see-saw masses. This SO(5) is the largest symmetry associated with 
neutrino mass terms and its implications are not yet fully explored.

\section{FlAVOR SYMMETRIES OF DENSE NEUTRINO GASES}

 Dense, self-interacting neutrino gases, encountered in astrophysical settings such as 
core-collapse supernovae, posses an interesting, non-linear SU(N$_{\rm flavor}$) 
symmetry. The standard MSW potential is provided by the coherent forward scattering of 
$\nu_e$'s off the  electrons in dense matter via W-exchange. There is a similar term 
with Z-exchange. But since it is the same for all neutrino flavors, it does not contribute 
at the tree level to phase differences that drive the MSW effect unless we invoke a sterile 
neutrino. Neutrinos play and salient role in the dynamics of core-collapse supernovae 
and hence it is crucial to explore all aspects of supernova neutrino physics  
\cite{Balantekin:2003ip}. 
If the neutrino density itself is also very high then one has to consider the effects 
of neutrinos scattering off other neutrinos; this is the case for a core-collapse supernova. 

For simplicity, let us consider only two flavors of neutrinos: electron neutrino, $\nu_e$, and 
another flavor, $\nu_x$. Introducing the creation and annihilation operators for one neutrino 
with three momentum ${\bf p}$, we can write down the generators of an SU(2) algebra 
\cite{Balantekin:2006tg}: 
\begin{eqnarray}
J_+({\bf p}) &=& a_x^\dagger({\bf p}) a_e({\bf p}), \> \> \>
J_-({\bf p})=a_e^\dagger({\bf p}) a_x({\bf p}), \nonumber \\
J_0({\bf p}) &=& \frac{1}{2}\left(a_x^\dagger({\bf p})a_x({\bf p})-a_e^\dagger({\bf p})a_e({\bf p})
\right). \label{su2}
\end{eqnarray}
Note that the integrals of these operators over all possible values of momenta also 
generate a global SU(2) algebra. 
Using the operators in Eq. (\ref{su2}) 
the Hamiltonian for a neutrino propagating through matter takes the form  
\begin{equation}
\label{msw}
 H_{\nu} = \int d^3{\bf p} \frac{\delta m^2}{2p} \left[
\cos{2\theta} J_0({\bf p}) + \frac{1}{2} \sin{2\theta}
\left(J_+({\bf p})+J_-({\bf p})\right) \right] -  \sqrt{2} G_F \int d^3{\bf p} 
\> N_e \>  J_0({\bf p}).  
\end{equation}
In Eq. (\ref{msw}), the first integral represents the neutrino mixing and the second integral 
represents the neutrino forward scattering off the background matter. Neutrino-neutrino 
interactions are described by the Hamiltonian 
\begin{equation}
\label{nunu}
H_{\nu \nu} = \sqrt{2} \frac{G_F}{V} \int d^3{\bf p} \> d^3{\bf q} \>  (1-\cos\vartheta_{\bf pq}) \> {\bf
J}({\bf p}) \cdot {\bf J}({\bf q}) ,
\end{equation}
where $\vartheta_{\bf pq}$ is the angle between neutrino momenta {\bf p} and {\bf q} and V 
is the normalization volume.  Note that the $(1-\cos\vartheta_{\bf pq}) $ term in the integral above 
comes from the V-A nature of the weak interactions and its presence is crucial to recover the 
effects of the weak 
interaction physics in the most general situation. Note that in the extremely idealized case of 
isotropic neutrino distribution and a very large number of neutrinos this term may average to a 
constant\footnote{Although the number of neutrinos in a core-collapse supernova is very large ($\sim 
10^{58}$), their distribution is very unlikely to be isotropic.} and the neutrino-neutrino interaction 
Hamiltonian simply reduces to the Casimir operator of the global SU(2) algebra. Inclusion of antineutrinos in Eqs. (\ref{msw}) and (\ref{nunu}) introduces a second set of SU(2) algebras. For three flavors one needs two sets of SU(3) algebras, one for neutrinos and one for antineutrinos 
\cite{Sawyer:2005jk}. 

Exact solutions of the combined Hamiltonian (Eqs. (\ref{msw}) and (\ref{nunu})) seem to be very 
difficult to obtain, but a saddle-point approximation \cite{Balantekin:2006tg} to its solutions, where neutrinos interact with a {\em neutrino} mean-field, is widely used in the literature to 
numerically investigate the underlying nonlinear behavior \cite{Qian:1994wh}. A recent comprehensive review is given in Ref. \cite{Duan:2009cd}. 

Unlike the typical many-body systems in condensed-matter physics (which are 
mostly controlled by electromagnetism) or in nuclear physics (which are 
mostly controlled by the strong force), this neutrino gas is the only example of a non-trivial 
many-body system entirely controlled by the weak interactions. 
This example illustrates that astrophysical extremes allow testing neutrino properties in ways 
that cannot be done elsewhere, e.g. $\nu-\nu$  effect as an Òemergent phenomenonÓ. 

\section{CP-VIOLATION AND NEUTRINO PHYSICS}

We now turn to the fundamental symmetry CP and its possible violation in the neutrino sector. 
The neutrino mixing matrix is 
\begin{equation}
 {\bf T}_{23}{\bf T}_{13}{\bf T}_{12}  = 
\left(
\begin{array}{ccc}
 1 & 0  & 0  \\
  0 & C_{23}   & S_{23}  \\
 0 & -S_{23}  & C_{23}  
\end{array}
\right)
\left(
\begin{array}{ccc}
 C_{13} & 0  & S_{13} e^{-i\delta}  \\
 0 & 1  & 0  \\
 - S_{13} e^{i \delta} & 0  & C_{13}  
\end{array}
\right) 
\left(
\begin{array}{ccc}
 C_{12} & S_{12}  & 0  \\
 - S_{12} & C_{12}  & 0  \\
0  & 0  & 1  
\end{array}
\right)
\end{equation}
where $C_{ij} = \cos \theta_{ij}$, $S_{ij} = \sin \theta_{ij}$, and $\delta$ is the CP-violating phase.  
Clearly a non-zero value of $\theta_{13}$ would also make the observation of the effects that 
depend on the CP-violating phase possible. There already are hints for a non-zero value 
of $\theta_{13}$ from solar, atmospheric, and reactor data \cite{Balantekin:2008zm,Fogli:2008jx}  
that will be probed by the reactor experiments in the near future 
\cite{Guo:2007ug,Ardellier:2006mn,Oh:2009zz}. 
 
It would be interesting to explore if matter amplifies or suppresses CP-violating effects. To this end 
introducing the operators \cite{Balantekin:1999dx} 
\[  \tilde{\Psi}_{\mu} = \cos \theta_{23} \Psi_{\mu} - \sin \theta_{23} \Psi_{\tau}, \]
\[  \tilde{\Psi}_{\tau} = \sin \theta_{23} \Psi_{\mu} + \cos \theta_{23} \Psi_{\tau}, \]
one can write down the neutrino evolution equations as 
\begin{equation} 
\label{CProt}
i \frac{\partial}{\partial t} 
\left(
\begin{array}{c}
  \Psi_e \\
 \tilde{ \Psi}_{\mu} \\
  \tilde{\Psi}_{\tau} 
\end{array}
\right) 
= \tilde{\bf H} 
\left(
\begin{array}{c}
  \Psi_e \\
  \tilde{\Psi}_{\mu} \\
  \tilde{\Psi}_{\tau} 
\end{array}
\right) 
\end{equation}
where 
\begin{equation}
\label{htilde}
\tilde{\bf H} = 
{\bf T}_{13}{\bf T}_{12} 
\left(
\begin{array}{ccc}
E_1  & 0  & 0  \\
0  & E_2  & 0  \\
0  &  0 & E_3  
\end{array}
\right) {\bf T}^{\dagger}_{12}{\bf T}^{\dagger}_{13}  + 
\left(   
\begin{array}{ccc}
 V_{e \mu} & 0  & 0  \\
 0 & S^2_{23}   V_{\tau \mu} & - C_{23} S_{23} V_{\tau \mu}  \\
0  & - C_{23} S_{23}V_{\tau \mu}   & C^2_{23} V_{\tau \mu}  
\end{array}
\right) .
\end{equation}
In writing Eq. (\ref{htilde}) a term proportional to identity is dropped by adding a term to all  
the matter potentials so that $V_{\mu \mu}=0$. 

If we can neglect the potential $V_{\tau \mu}$ it is straightforward to show that 
\[
\tilde{H} (\delta) = {\bf S} \tilde{H} (\delta=0) {\bf S}^{\dagger}
\]
with
\[
{\bf S} = \left(
\begin{array}{ccc}
 1 & 0  & 0  \\
 0 & 1  & 0  \\
 0 & 0  & e^{i \delta}  
\end{array}
\right) .
\] 
This factorization gives us interesting sum rules:
 Electron neutrino survival probability, $P (\nu_e \rightarrow \nu_e)$ is independent of the value of the CP-violating phase, $\delta$; or equivalently
the combination $P (\nu_{\mu} \rightarrow \nu_e) + P (\nu_{\tau} \rightarrow \nu_e)$ at a fixed energy is independent of the value of the CP-violating phase \cite{Balantekin:2007es}. 
 It is possible to derive similar sum rules for other amplitudes \cite{Kneller:2009vd}. 
These results hold even if the neutrino-neutrino interactions are included in the Hamiltonian 
\cite{Gava:2008rp}.
Electron neutrino flux at a distance $r$ from the neutrinosphere is
\[
\mathcal{L}^{(r)} _e = \mathcal{L}^{(0)}_e P(\nu_e \rightarrow \nu_e, r) 
+  \mathcal{L}^{(0)}_{\mu}  P(\nu_{\mu} \rightarrow \nu_e, r)  
+  \mathcal{L}^{(0)}_{\tau} P(\nu_{\tau} \rightarrow \nu_e, r) .
\]
If the $\nu_{\mu}$ and $\nu_{\tau}$ luminosities are the same at the neutrinosphere, i.e. 
$  \mathcal{L}^{(0)}_{\mu}  =   \mathcal{L}^{(0)}_{\tau}$, we get 
\[
\mathcal{L}^{(r)} _e = \mathcal{L}^{(0)}_e \left\{ P(\nu_e \rightarrow \nu_e, r) \right\} 
+  \mathcal{L}^{(0)}_{\mu} \left\{  P(\nu_{\mu} \rightarrow \nu_e, r)  
+   P(\nu_{\tau} \rightarrow \nu_e, r) \right\}
\]
Since in the factorizable  limit the quantities inside the curly brackets do not depend on the CP-violating phase, $\delta$, we conclude that
\[
\mathcal{L}^{(r)} _e ( \delta \neq 0) = \mathcal{L}^{(r)} _e ( \delta = 0) , 
\]
i.e. under the assumptions stated above, electron neutrino survival probability and, consequently, electron neutrino and antineutrino luminosities are independent of the CP-violating phase. To be able to observe the effects of $\delta$, we need to relax the underlying assumptions by either i) Permitting  the $\nu_{\mu}$ and 
$\nu_{\tau}$ luminosities to be different at the neutrinosphere (Standard Model (SM)  loop corrections and also physics beyond the Standard Model may do this); or ii) Exploring when $V_{\tau \mu}$ is non-zero due to SM loop corrections or due to physics beyond SM.  The impact of such loop corrections on supernova physics was first explored in Ref. \cite{Akhmedov:2002zj}.

This work was supported in part
by the U.S. National Science Foundation Grant No. PHY-0855082 
and
in part by the University of Wisconsin Research Committee with funds
granted by the Wisconsin Alumni Research Foundation.

\end{document}